\documentclass[lettersize,journal]{IEEEtran}
\usepackage{amsmath,,amssymb,amsfonts}
\usepackage{algorithmic}
\usepackage{algorithm}
\usepackage{array}
\usepackage{color}
\usepackage{textcomp}
\usepackage{stfloats}
\usepackage{url}
\usepackage{bm}
\usepackage{verbatim}
\usepackage{graphicx}
\usepackage{cite}
\usepackage{subfigure} 
\usepackage{amsthm}
\usepackage{caption}
\usepackage{graphicx}
\usepackage{float} 
\usepackage{subcaption}
\begin{document}

\title{Cooperative Double IRS aided Secure Communication for MIMO-OFDM Systems}

\author{Weijie Xiong, Jingran Lin, Di Jiang, Yuhan Zhang, Kai Zhong, Qiang Li, and Cunhua Pan
\thanks{This work was supported in part by the Natural Science Foundation of China (NSFC) under Grants 62171110 and 62501112, and in part by the China Postdoctoral Science Foundation under Grant 2025M773511. \textit{(Corresponding author: Jingran Lin, Di Jiang)}. Jingran Lin and Qiang Li are with University of Electronic Science and Technology of China, Chengdu 611731, China, the Laboratory of Electromagnetic Space Cognition and Intelligent Control, Beijing 100083, China, and also with the Tianfu Jiangxi Laboratory, Chengdu, Sichuan 641419, China (e-mail: jingranlin@uestc.edu.cn; lq@uestc.edu.cn). Weijie Xiong, Di Jiang, Yuhan Zhang, Kai Zhong are with University of Electronic Science and Technology of China, Chengdu 611731, China (e-mail: 202311012313@std.uestc.edu.cn; dijiang@uestc.edu.cn; 202412012225@std.uestc.edu.cn; 201921011206@std.uestc.edu.cn). Cunhua Pan is with the National Mobile Communications Research Laboratory, Southeast University, China (e-mail: cpan@seu.edu.cn).}
}

\markboth{Journal of \LaTeX\ Class Files,~Vol.~14, No.~8, August~2021}%
{Shell \MakeLowercase{\textit{et al.}}: A Sample Article Using IEEEtran.cls for IEEE Journals}


\maketitle

\begin{abstract}
Cooperative double intelligent reflecting surface (double-IRS) has emerged as a promising approach for enhancing physical layer security (PLS) in MIMO systems. However, existing studies are limited to narrowband scenarios and fail to address wideband MIMO-OFDM. In this regime, frequency-flat IRS phases and cascaded IRS links cause severe coupling, rendering narrowband designs inapplicable. To overcome this challenge, we introduce cooperative double-IRS-assisted wideband MIMO-OFDM and propose an efficient manifold-based solution. By regarding the power and constant modulus constraints as Riemannian manifolds, we reformulate the non-convex secrecy sum rate maximization as an unconstrained optimization on a product manifold. Building on this formulation, we further develop a product Riemannian gradient descent (PRGD) algorithm with guaranteed stationary convergence. Simulation results demonstrate that the proposed scheme effectively resolves the OFDM coupling issue and achieves significant secrecy rate gains, outperforming single-IRS and distributed multi-IRS benchmarks by 32.0\% and 22.3\%, respectively.
\end{abstract}


\section{Introduction}
Orthogonal frequency division multiplexing (OFDM) has been widely adopted in modern wireless systems, including 5G New Radio and IEEE 802.11, for its spectral efficiency and implementation flexibility \cite{deng2025unifying}. As a wideband and frequency selective physical layer, OFDM partitions the spectrum into subcarriers that experience distinct channels. When combined with multiple input multiple output (MIMO) antenna arrays, it enables beamforming, spatial multiplexing, and waveform diversity, yielding substantial performance gains \cite{wang2024tutorial}. Consequently, MIMO-OFDM has become a key technology in contemporary wireless networks \cite{li2023cell}.

Despite its advantages, the broadcast nature and spectrum sharing of MIMO-OFDM expose it to security threats. Cryptographic methods can ensure confidentiality but incur heavy computational overhead and complex key management \cite{khoshafa2024ris}. These drawbacks have motivated interest in physical layer security (PLS) \cite{xiong2025constant}. Existing PLS techniques, such as artificial noise \cite{nizam2023enhancing}, cooperative jamming \cite{jiang2024improving}, secure beamforming \cite{xiong2025secure,11276865}, and directional modulation \cite{chen2024physical}, enhance secrecy but become ineffective when the eavesdropping channel is highly correlated with the legitimate user’s channel \cite{rafieifar2023secure}.

Intelligent reflecting surfaces (IRSs) have emerged as a promising solution to reconfigure wireless propagation through programmable phase shifts, thus creating controllable paths that decouple legitimate and eavesdropping channels \cite{rafieifar2023secure}. This makes IRSs an efficient means of enhancing PLS with clear advantages over conventional approaches \cite{kaur2024survey}. However, in complex MIMO-OFDM environments with numerous obstacles, single-IRS or distributed multi-IRS deployments are often insufficient, since secure transmission often relies on a single-reflection path that lacks spatial diversity and adaptability, leaving it highly vulnerable to eavesdropping \cite{zhou2024secure,cao2022cooperative}.

To overcome these shortcomings, cooperative IRSs have been proposed, where multiple IRSs jointly reflect signals toward the legitimate user through coordinated phase-shift design \cite{cao2022cooperative}. By offering richer spatial diversity and greater flexibility in controlling the propagation environment, cooperative IRSs can effectively mitigate multipath fading and bypass obstacles. Recent studies have demonstrated their potential to significantly enhance PLS in SISO \cite{zhou2024secure}, MISO \cite{cao2022cooperative,xu2024robust,jiang2021joint}, and MIMO systems \cite{wang2024joint}, making them a promising solution for secure wireless transmission. Nevertheless, most of these works focus on narrowband single-carrier models, leaving open the question of whether similar gains can be achieved in broadband MIMO-OFDM systems.

To bridge this gap, this paper tackles the joint design of transmit beamforming and IRS reflection coefficients for a cooperative double-IRS-assisted MIMO-OFDM system. The key challenges stem from two coupling effects: 1) cross-subcarrier coupling due to the shared frequency-flat IRS phases, and 2) multiplicative coupling between the two IRS phase vectors via the cascaded link. These effects render conventional narrowband methods inapplicable \cite{zhou2024secure,cao2022cooperative,xu2024robust,jiang2021joint,wang2024joint}. To this end, we first formulate the secrecy sum rate (SSR) maximization problem, and then, by exploiting the manifold structure of the constraints, we develop a product Riemannian gradient descent (PRGD) algorithm with guaranteed convergence to solve it. Simulations show that the proposed scheme effectively resolves the coupling issues and achieves consistent secrecy gains over single-IRS and distributed multi-IRS baselines.

\section{System Model and Problem Formulation}
\label{sec:format}

\begin{figure}[htbp]
  \begin{center}
  \includegraphics[width=2.5in]{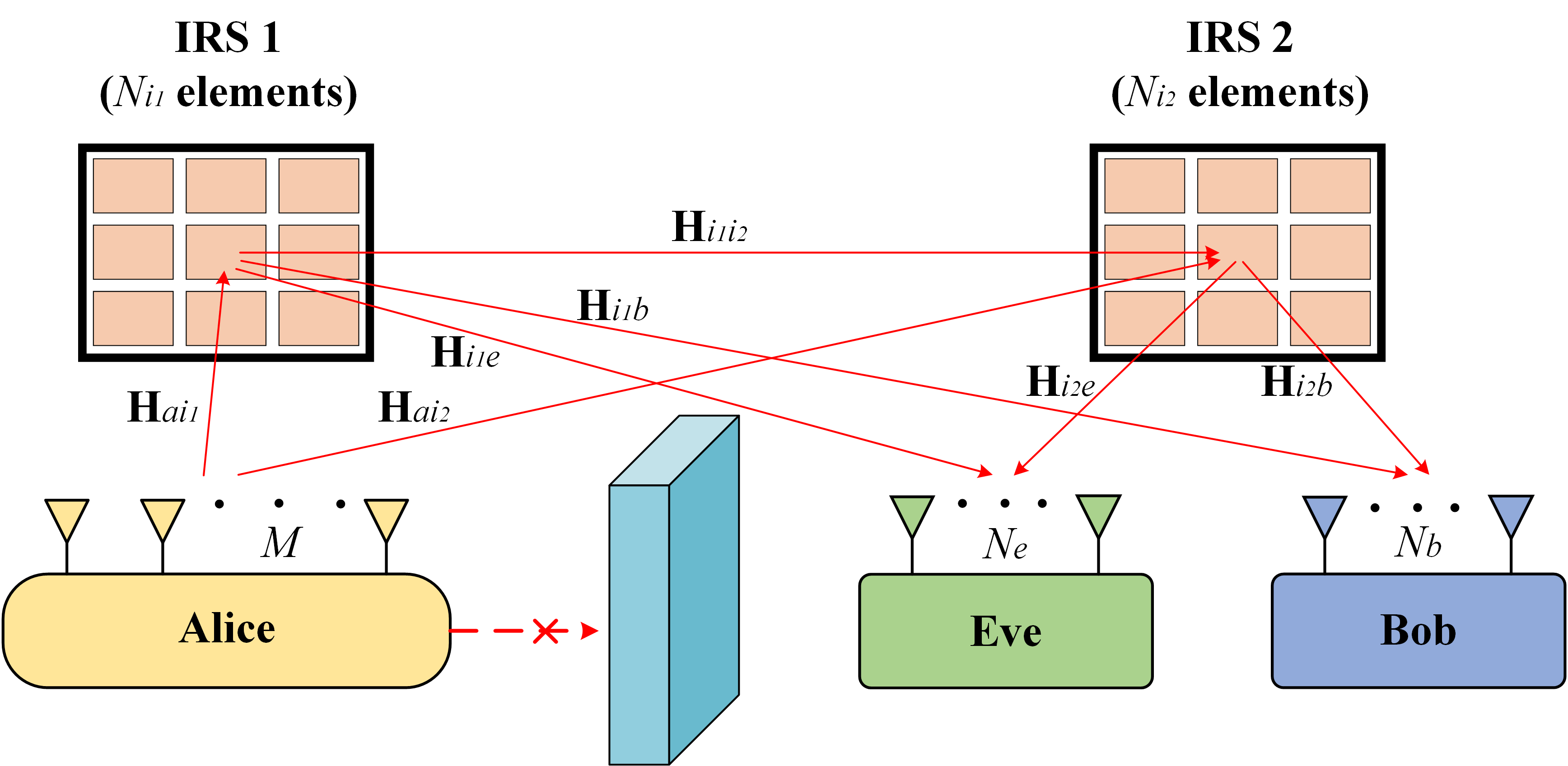}\\
  \caption{\footnotesize MIMO-OFDM system illustration with cooperative double-IRS.}\label{1}
  \end{center}
  \vspace{-0.5cm}
\end{figure}

We consider a cooperative double-IRS-aided MIMO-OFDM system, as shown in Fig. \ref{1}, where the transmitter (Alice) equipped with \(M\) antennas broadcasts common confidential information to a legitimate user (Bob) with \(N_b\) antennas, in the presence of an eavesdropper (Eve) equipped with \(N_e\) antennas. Due to the blocked direct link between Alice and Bob, two IRSs (IRS 1 and IRS 2), deployed near Alice and Bob, are employed to cooperatively assist secure communication, where IRS 1 and IRS 2 comprise \(N_{i_1}\) and \(N_{i_2}\) reflecting elements (REs), respectively.

Define the number of OFDM subcarriers as $K$, ${\bf H}_{a{i_1},k} \in \mathbb{C}^{M \times N_{i_1}}$, ${\bf H}_{a{i_2},k} \in \mathbb{C}^{M \times N_{i_2}}$, ${\bf H}_{{{i_1}b},k} \in \mathbb{C}^{N_{i_1} \times N_b}$, ${\bf H}_{{{i_1}e},k} \in \mathbb{C}^{N_{i_1} \times N_e}$, ${\bf H}_{{{i_2}b},k} \in \mathbb{C}^{N_{i_2} \times N_b}$, ${\bf H}_{{{i_2}e},k} \in \mathbb{C}^{N_{i_2} \times N_e}$, and ${\bf H}_{{i_1}{i_2},k} \in \mathbb{C}^{N_{i_1} \times N_{i_2}}$ stand for the complex baseband channels from Alice to IRS 1, Alice to IRS 2, IRS 1 to Bob, IRS 1 to Eve, IRS 2 to Bob, IRS 2 to Eve, and IRS 1 to IRS 2 on the $k$-th subcarrier respectively. Due to practical hardware constraints, the IRS can only apply identical phase shifts across all subcarriers. Thus, we define phase shifts for IRS 1 and IRS 2 on any subcarrier as ${\bm \phi}_1 = [\phi_{1,1},...,\phi_{1,n_{i_1}},...,\phi_{1,N_{i_1}}]\in \mathbb{C}^{N_{i_1}},\forall n_{i_1} \in [1,N_{i_1}]$ and ${\bm \phi}_2 = [\phi_{2,1},...,\phi_{2,n_{i_2}},...,\phi_{2,N_{i_2}}]\in \mathbb{C}^{N_{i_2}},\forall n_{i_2} \in [1,N_{i_2}]$, respectively.

Let ${\bf s}_k \in \mathbb{C}^{N_s}$ denote the coded confidential information
for Bob on the $k$-th subcarrier, where $N_s$ represents the length of the data streams and the corresponding beamforming vector is ${\bf W}_k \in \mathbb{C}^{M \times N_s}$. Then, the signal received by Bob and Eve on the $k$-th subcarrier is respectively expressed by\footnote{We assume that Eve’s channel state information (CSI) is available at Alice. This assumption is reasonable when Eve is an untrusted internal user, an active Eve whose channel is estimable from its signals, or a passive Eve whose channel can be inferred from local-oscillator (LO) leakage \cite{10922160}.}\textsuperscript{,}\footnote{We assume perfect CSI to explore the performance limits of the system. While we explicitly acknowledge that acquiring separated channel matrices is practically challenging for passive IRSs where typically only cascaded channels are estimated \cite{zheng2021efficient}, this assumption allows us to establish a theoretical upper bound for the cooperative architecture. Furthermore, to address the inevitable uncertainties in any practical estimation scenario, we evaluate the scheme's robustness against channel estimation errors (CEEs) in Fig. \ref{NMSE}.},
\begin{equation}
\begin{aligned}
    & {\bf y}_{b,k} = {\bf H}_{b,k}{\bf W}_k{\bf s}_k + {\bf n}_{b,k},\quad \forall k \in K,\\
    &{\bf y}_{e,k} = {\bf H}_{e,k}{\bf W}_k{\bf s}_k + {\bf n}_{e,k},\quad \forall k \in K,  \label{recieveing}
\end{aligned}    
\end{equation}
where we assume that all subcarriers are subject to the same additive white Gaussian noise for both Bob and Eve, modeled as ${\bf{n}}_{b,k} \sim \mathcal{CN}(\mathbf{0}, \sigma_b^2 {\bf I}_{N_b}), \ \forall k$, and ${\bf{n}}_{e,k} \sim \mathcal{CN}(\mathbf{0}, \sigma_e^2 {\bf I}_{N_e}), \ \forall k$, respectively, with $\mathbf{I}$ denoting the identity matrix, and,
\begin{equation}
\begin{aligned}
        & {\bf H}_{b,k}= \left\{\begin{array}{l} {\bf H}_{ai_1,k}\text{diag}({\bm \phi}_1){\bf H}_{i_1b,k}\\+{\bf H}_{ai_2,k}\text{diag}({\bm \phi}_2){\bf H}_{i_2b,k}\\+{\bf H}_{ai_1,k}\text{diag}({\bm \phi}_1){\bf H}_{i_1i_2,k}({\bm \phi}_2){\bf H}_{i_2b,k} \end{array}\right\} ,\\
    &{\bf H}_{e,k}=\left\{\begin{array}{l}{\bf H}_{ai_1,k}\text{diag}({\bm \phi}_1){\bf H}_{i_1e,k}\\+{\bf H}_{ai_2,k}\text{diag}({\bm \phi}_2){\bf H}_{i_2e,k}\\+{\bf H}_{ai_1,k}\text{diag}({\bm \phi}_1){\bf H}_{i_1i_2,k}({\bm \phi}_2){\bf H}_{i_2e,k}\end{array}\right\}.  \label{recieveingchannel}
\end{aligned}    
\end{equation}

Based on (\ref{recieveing}) and (\ref{recieveingchannel}), the SSR from Alice to Bob for $k$-th subcarrier is given by \cite{wang2024joint},
\begin{equation}
R_{sec,k} = [R_{b,k} - R_{e,k}]^+, \quad \forall k \in K,
\end{equation} 
where $[\cdot]^+ =\max(0,\cdot)$, and \( R_{b,k} \) and \( R_{e,k} \) denote the achievable rates of Bob’s and Eve’s links on the $k$-th subcarrier, respectively, given by,
\begin{equation}
\begin{aligned}
        & R_{b,k}=\log \text{det}( {\bf I}_{N_b} + \tfrac{1}{\sigma_b^2}{\bf H}_{b,k}{\bf W}_k{\bf W}_k^H{\bf H}_{b,k}^H ) ,\\
     & R_{e,k}=\log \text{det}({\bf I}_{N_e} + \tfrac{1}{\sigma_e^2} {\bf H}_{e,k}{\bf W}_k{\bf W}_k^H{\bf H}_{e,k}^H ),  \label{reciveSR}
\end{aligned}    
\end{equation}
where $\text{det}(\cdot)$ and $(\cdot)^H$ denote the determinant and conjugate transpose operators, respectively.

In this paper, we aim to maximize the SSR across all subcarriers by jointly optimizing the transmit beamforming \( \{\mathbf{W}_k\}_{k=1}^K \) at Alice and the phase shifts of IRS 1 (\( {\bm \phi}_1 \)) and IRS 2 (\( {\bm \phi}_2 \)), subject to the total transmit power constraint and the structural constraints of the IRSs. Accordingly, the SSR maximization problem is formulated as follows {\footnote{Without the IRS 1 to IRS 2 channel ${\bf H}_{i_1i_2}$, the cooperative double-IRS design reduces to a single-IRS or a distributed multi-IRS system, in which the objective function can be reformulated by merging the variables ${\bm \phi}_1$ and ${\bm \phi}_2$ into a single variable \cite{zhou2024secure,cao2022cooperative}. However, when ${\bf H}_{i_1i_2}$ is present, such merging becomes infeasible due to the additional terms ${\bf H}_{ai_1,k}\text{diag}({\bm \phi}_1){\bf H}_{i_1i_2,k}\text{diag}({\bm \phi}_2){\bf H}_{i_2b,k}$ (and $\mathbf{H}_{i2e,k}$ for Eve). These terms increase the optimization complexity but also provide additional spatial diversity and greater flexibility in controlling the propagation environment.}\textsuperscript{,}}{\footnote{Compared with narrowband single-carrier designs that assume $K=1$ \cite{zhou2024secure,cao2022cooperative,xu2024robust,jiang2021joint,wang2024joint}, the wideband MIMO-OFDM case ($K>1$) introduces two additional couplings. First, the frequency-flat IRS phase vectors $\boldsymbol{\phi}_1$ and $\boldsymbol{\phi}_2$ are shared across all subcarriers, which non-separably couples the per-subcarrier beamformers $\{\mathbf{W}_k\}_{k=1}^{K}$ through $\mathbf{H}_{b,k}$ and $\mathbf{H}_{e,k}$. Second, the IRS-to-IRS cascade $\mathbf{H}_{ai1,k}\operatorname{diag}(\boldsymbol{\phi}_1)\mathbf{H}_{i1i2,k}\operatorname{diag}(\boldsymbol{\phi}_2)\mathbf{H}_{i2b,k}$ (and $\mathbf{H}_{i2e,k}$ for Eve), induces multiplicative coupling between $\boldsymbol{\phi}_1$ and $\boldsymbol{\phi}_2$ on each subcarrier. These couplings yield a joint optimization problem over all subcarriers and variables, rendering narrowband decomposition methods inapplicable.}},
\begin{subequations}
\begin{align}
 \max _{\{{\bf W}_k\}_{k=1}^K, {\bm \phi}_1,{\bm \phi}_2} &\quad \sum_{k=1}^K R_{sec,k}, \label{objorig}\\
 \text { s.t. }&\quad \sum_{k=1}^K \left\|{\bf W}_k\right\|_F^2\le P, \label{cons1orig}\\
       &\quad |{\phi}_{1,n_1}| = 1,\forall n_1 \in N_{i_1}, \label{endorig}\\
       &\quad |{\phi}_{2,n_2}| = 1,\forall n_2 \in N_{i_2}, \label{endorig2}
\end{align}
\label{overallproblem}%
\end{subequations}
where $P_k$ is the transmit power on the \(k\)-th subcarrier, while $\|\cdot\|_F$ and $|\cdot|$ denote the Frobenius norm and the absolute value operator, respectively. Generally, problem (\ref{overallproblem}) is challenging to solve due to the following factors: the non-concave objective function (\ref{objorig}), the coupling effect between the variables $\{{\bf W}_k\}_{k=1}^K$, ${\bm \phi}_1$, and ${\bm \phi}_2$ in the objective function (\ref{objorig}), and the non-convex constraints in (\ref{endorig}) and (\ref{endorig2}).

\section{Problem Reformulation and Proposed Method}
In light of the above challenges, we propose a tractable approach to solve problem (\ref{overallproblem}). Although alternating optimization (AO) \cite{zhang2021multiple} is commonly used for highly coupled variables, it typically lacks convergence guarantees to a stationary point for the overall problem. To overcome this, we reformulate problem (\ref{overallproblem}) as an unconstrained optimization on a product Riemannian manifold that jointly encompasses all variables. We then develop an efficient product Riemannian gradient descent (PRGD) algorithm that provably converges to a stationary point. The details are provided below.

\subsection{Problem Reformulation}
To begin with, since the maximum SSR is achieved when the total transmit power is fully utilized, i.e., $\sum_{k=1}^K \left\|{\bf W}_k\right\|_F^2= P$, the constraints in (\ref{cons1orig}) can be equivalently reformulated as,
\begin{equation}
\begin{aligned}
    \sum_{k=1}^K \left\|{\bf W}_k\right\|_F^2 = P& \Leftrightarrow  \text{Tr}([{\bf W}_1;...;{\bf W}_K][{\bf W}_1;...;{\bf W}_K]^H) =P\\ &  \Leftrightarrow \text{Tr}({\bf \tilde  W}{\bf \tilde W}^H)=P\Leftrightarrow \|{\bf \tilde W}\|_F = \sqrt{P},
\end{aligned}    
\end{equation}
where $\text{Tr}(\cdot)$ denotes the trace operator, and ${\bf \tilde W}=[{\bf W}_1;...;{\bf W}_K] \in \mathbb{C}^{M\times KN_s}$. Furthermore, secure communication necessitates a positive SSR in problem (\ref{overallproblem}). Based on this observation, by interchanging the numerator and denominator in (\ref{objorig}), problem (\ref{overallproblem}) can be reformulated as follows,
\begin{subequations}
\begin{align}
 \min _{{\bf \hat W}, {\bm \phi}_1,{\bm \phi}_2} &\quad \sum_{k=1}^K \log\frac{\text{det}({\bf I}_{N_e} + \tfrac{P}{\sigma_e^2} {\bf H}_{e,k}{\bf W}_k{\bf W}_k^H{\bf H}_{e,k}^H )}{\text{det}( {\bf I}_{N_b} + \tfrac{P}{\sigma_b^2}{\bf H}_{b,k}{\bf W}_k{\bf W}_k^H{\bf H}_{b,k}^H )}, \label{objorig2}\\
 \text { s.t. }&\quad \|{\bf \hat W}\|_F = 1, \label{cons1orig2}\\
       &\quad |{\phi}_{1,n_1}| = 1,\forall n_1 \in N_{i_1}, \label{endorignew}\\
       &\quad |{\phi}_{2,n_2}| = 1,\forall n_2 \in N_{i_2}, \label{endorig2new}
\end{align}
\label{overallproblem2}%
\end{subequations}
where we normalize the Frobenius norm of ${\bf\tilde W}$ to unity by introducing ${\bf\hat W}={\bf\tilde W}/\sqrt{P}$ and scaling the channels by $\sqrt{P}$. It can be observed that constraint (\ref{cons1orig2}) defines a complex sphere manifold and constraints (\ref{endorignew}) and (\ref{endorig2new}) describe complex circle manifolds within Riemannian space, defined as \cite{boumal2023intromanifolds}, 
\begin{equation}
\begin{aligned}
&\mathcal{M}_{\bf \hat W}=\left\{{\bf \hat W} \in \mathbb{C}^{M\times KN_s} \mid \|{\bf \hat W}\|_F=1 \right\},\\& \mathcal{M}_{{\bm \phi}_1}=\left\{{\bm \phi}_1 \in \mathbb{C}^{N_{i_1} } \mid |{\phi}_{n_{i_1}}|=1,\forall n_{i_1} \right\},\\&\mathcal{M}_{{\bm \phi}_2}=\left\{{\bm \phi}_2 \in \mathbb{C}^{N_{i_2} } \mid |{\phi}_{n_{i_2}}|=1,\forall n_{i_2} \right\}.
\end{aligned}
\end{equation}
This observation motivates us to address problem (\ref{overallproblem2}) using manifold-based optimization techniques \cite{xiong2025enhancing}. However, the coupling between variables poses a challenge to directly applying these methods. To overcome this, as demonstrated in \cite{boumal2023intromanifolds}, multiple Riemannian manifolds can be combined into a single entity known as a product Riemannian manifold (PRM). This approach integrates the characteristics of the individual manifolds and effectively mitigates variable coupling. Building upon this concept, we combine \(\mathcal{M}_{\bf \hat W}\), \(\mathcal{M}_{{\bm \phi}_1}\), and \(\mathcal{M}_{{\bm \phi}_2}\) to form the PRM \(\mathcal{M}\), defined as follows,
\begin{equation}
\begin{aligned}
&\mathcal{M} = \mathcal{M}_{\bf \hat W} \times \mathcal{M}_{{\bm \phi}_1} \times \mathcal{M}_{{\bm \phi}_2} \\&= \left\{({\bf \hat W}, {\bm \phi}_1, {\bm \phi}_2): {\bf \hat W} \in \mathcal{M}_{{\bf \hat W}}, {\bm \phi}_1 \in \mathcal{M}_{{\bm \phi}_1} ,{\bm \phi}_1 \in \mathcal{M}_{{\bm \phi}_2}\right\},
\end{aligned}
\end{equation}
where $({\bf \hat W}, {\bm \phi}_1, {\bm \phi}_2)$ denotes the collection of the variables set. By constructing PRM, we are able to reformulate (\ref{overallproblem2}) into an unconstrained optimization problem over Riemannian space, given as,
\begin{equation}
\min _{{\bm \Upsilon}\in\mathcal{M}} f({\bm \Upsilon}), \label{consr}
\end{equation}
where we use ${\bm \Upsilon} = {\bf \hat W} \oplus {\bm \phi}_1 \oplus {\bm \phi}_2$ to denote the variable set $({\bf \hat W}, {\bm \phi}_1, {\bm \phi}_2)$ for brevity, with $\oplus$ representing the direct-sum operator defined as ${\bf A} \oplus {\bf B} = \mathrm{blkdiag}({\bf A}, {\bf B})$. To solve the problem (\ref{consr}) efficiently, in the following, a highly efficient PRGD algorithm is proposed. The details are given as below.

\subsection{Proposed PRGD Algorithm}
Unlike gradient descent in Euclidean space, applying it directly on curved spaces like Riemannian manifolds is challenging due to their non-Euclidean geometry. To enable optimization, we construct a tangent space as a local linear approximation of the manifold, allowing gradient computation and descent steps in a Euclidean-like setting. This approach preserves the geometric constraints of the original problem. Specifically, the tangent space \( \mathcal{T}_{\bm \Upsilon}\mathcal{M} \) of PRM \( \mathcal{M} \) is defined as \cite{boumal2023intromanifolds},
\begin{equation}
{\mathcal T}_{\bf \Upsilon}{\mathcal M} = {\mathcal T}_{\bf \hat W}\mathcal{M}_{\bf \hat W} \oplus {\mathcal T}_{{\bm \phi}_1}\mathcal{M}_{{\bm \phi}_1} \oplus  {\mathcal T}_{{\bm \phi}_2}\mathcal{M}_{{\bm \phi}_2}, \label{tanspcae}
\end{equation}
where \( {\mathcal T}_{\bf \hat W}\mathcal{M}_{\bf \hat W} \), \( {\mathcal T}_{{\bm \phi}_1}\mathcal{M}_{{\bm \phi}_1} \), and \( {\mathcal T}_{{\bm \phi}_2}\mathcal{M}_{{\bm \phi}_2} \) are the tangent spaces of the individual manifolds \( \mathcal{M}_{\bf \hat W} \), \( \mathcal{M}_{{\bm \phi}_1} \), and \( \mathcal{M}_{{\bm \phi}_2} \), respectively, defined as,
\begin{equation}
    \begin{aligned}
       & {\mathcal T}_{\bf \hat W}\mathcal{M}_{\bf \hat W} = \{ {\bf \Xi}\in\mathbb{C}^{M \times KN_s} \mid \Re\{\text{Tr}\{{\bf \Xi}^H{\bf W}\}\}= 0 \},\\
       & {\mathcal T}_{{\bm \phi}_1}\mathcal{M}_{{\bm \phi}_1} = \{ {\bm \psi}_1\in\mathbb{C}^{N_{i_1}} \mid \Re\{{\bm \psi}_1^*  \odot {\bm \phi}_1\}= 0 \},\\
       & {\mathcal T}_{{\bm \phi}_2}\mathcal{M}_{{\bm \phi}_2} = \{ {\bm \psi}_2\in\mathbb{C}^{N_{i_2}} \mid \Re\{{\bm \psi}_2^*  \odot {\bm \phi}_2\}= 0 \},
    \end{aligned}
\end{equation}
where ${\bf \Xi}$, ${\bm \psi}_1$, and ${\bm \psi}_2$ are tangent vectors within their respective tangent spaces, $\Re(\cdot)$, $(\cdot)^*$, and $\odot$ denote the real part, conjugate, and Hadamard product operators, respectively.

Benefiting from the Euclidean-like structure of the tangent space \( \mathcal{T}_{\bf \Upsilon} \mathcal{M} \), gradient descent can be performed within it \cite{boumal2023intromanifolds}. However, since updates in \( \mathcal{T}_{\bf \Upsilon} \mathcal{M} \) may lead to points outside the manifold, a retraction operation is required to project the updated point back onto \( \mathcal{M} \). Based on the above discussion, the main steps of the PRGD algorithm for solving problem (\ref{consr}) at $q$-th iteration consists of three steps \cite{boumal2023intromanifolds}: 1) computing the product Riemannian gradient of the objective function, 2) selecting a suitable step size, and 3) updating the iterate followed by a retraction. These steps are repeated until convergence.

\subsubsection{Calculation of the Product Riemannian Gradient}
To perform gradient descent over the tangent space \( \mathcal{T}_{\bf \Upsilon} \mathcal{M} \), a descent direction is required. Similar to the Euclidean case, where the negative Euclidean gradient is used as the descent direction, in the context of Riemannian manifolds, the descent direction is given by the negative Riemannian gradient \cite{boumal2023intromanifolds}. The Riemannian gradient $\text{grad}_{\mathcal{M}} f({\bf \Upsilon})$ is obtained by projecting the Euclidean gradient onto the tangent space, as follows,
\begin{equation}
\begin{aligned}
      & \text{grad}_{\mathcal{M}} f({\bf \Upsilon})= \left\{\begin{array}{l}  \text{grad}_{\mathcal{M}_{\bf \hat W}}f({\bf \Upsilon})\\\oplus\text{grad}_{\mathcal{M}_{{\bm \phi}_1}}f({\bf \Upsilon}) \\\oplus \text{grad}_{\mathcal{M}_{{\bm \phi}_2}}f({\bf \Upsilon})\end{array} \right\}  \\ &=\left\{\begin{array}{l}  (\nabla_{{\bf \hat W}} f({\bf \Upsilon})-\Re\{\text{Tr}\{\nabla_{{\bf \hat W}}^H f({\bf \Upsilon}){\bf \hat W}\}\}{\bf \hat W})\\ 
      \oplus(\nabla_{{\bm \phi}_1} f({\bf \Upsilon})-\Re\{\nabla^*_{{\bm \phi}_1}f({\bf \Upsilon})\odot {\bm \phi}_1 \}\odot{\bm \phi}_1) \\\oplus(\nabla_{{\bm \phi}_2} f({\bf \Upsilon})-\Re\{\nabla^*_{{\bm \phi}_2}f({\bf \Upsilon})\odot {\bm \phi}_2 \}\odot{\bm \phi}_2)\end{array} \right\},\label{calculReg}
\end{aligned}       
\end{equation}
where \( \text{grad}_{\mathcal{M}_{\bf \hat W}}f({\bf \Upsilon}) \), \( \text{grad}_{\mathcal{M}_{{\bm \phi}_1}}f({\bf \Upsilon}) \), and \( \text{grad}_{\mathcal{M}_{{\bm \phi}_2}}f({\bf \Upsilon}) \) are the Riemannian gradients associated with the individual manifolds \( \mathcal{M}_{\bf \hat W} \), \( \mathcal{M}_{{\bm \phi}_1} \), and \( \mathcal{M}_{{\bm \phi}_2} \), respectively. The terms $\Re\{\text{Tr}(\nabla_{{\bf \hat W}}^H f({\bf \Upsilon}){\bf \hat W})\}{\bf \hat W}$, $\Re\{\nabla^*_{{\bm \phi}_1}f({\bf \Upsilon}) \odot {\bm \phi}_1\} \odot {\bm \phi}_1$, and $\Re\{\nabla^*_{{\bm \phi}_2}f({\bf \Upsilon}) \odot {\bm \phi}_2\} \odot {\bm \phi}_2$ denote the orthogonal projections of the Euclidean gradients onto their respective manifolds, which are subtracted to ensure that the Riemannian gradients remain in the corresponding tangent spaces. Besides, \( \nabla_{{\bf \hat W}} f({\bf \Upsilon}) \), \( \nabla_{{\bm \phi}_1} f({\bf \Upsilon}) \), and \( \nabla_{{\bm \phi}_2} f({\bf \Upsilon}) \) denote the Euclidean gradients, given by (\ref{eua1}), (\ref{eua2}), and (\ref{eua3}), respectively,
\begin{subequations}
\begin{align}
    &\nabla_{{\bf \hat W}} f({\bf \Upsilon})=[\nabla_{{\bf W}_1} f({\bf \Upsilon});...;\nabla_{{\bf W}_K} f({\bf \Upsilon})],\notag\\
    &\nabla_{{\bf W}_k} f({\bf \Upsilon}) =  2({\bf H}_{e,k}^H{\bf Q}_{e,k}-{\bf H}_{b,k}^H{\bf Q}_{b,k}){\bf W}_k,\forall k,\label{eua1}\\
    &\nabla_{{\bm \phi}_1}f({\bf \Upsilon})= 2\textstyle\sum_{k=1}^K\text{diag}(({\bf H}_{{i_1}e,k}^H{\bf Q}_{e,k}-{\bf H}_{{i_1}b,k}^H{\bf Q}_{b,k}\notag\\&+{\bf H}_{{i_1}{i_2},k}^H\text{diag}^H({\bm \phi}_2)({\bf H}_{{i_2}e,k}^H{\bf Q}_{e,k}-{\bf H}_{{i_2}b,k}^H{\bf Q}_{b,k})){\bf M}_{i_1,k}),\label{eua2}\\
    &\nabla_{{\bm \phi}_2}f({\bf \Upsilon})= 2\textstyle\sum_{k=1}^K\text{diag}(({\bf H}_{{i_2}e,k}^H{\bf Q}_{e,k}-{\bf H}_{{i_2}b,k}^H{\bf Q}_{b,k})\notag\\&\quad\quad\quad\quad\quad\quad({\bf M}_{i_2,k}+{\bf M}_{i_1,k}\text{diag}^H({\bm \phi}_1){\bf H}_{{i_1}{i_2},k}^H)),\label{eua3}
\end{align}    
\end{subequations}    
where,
\begin{equation}
\begin{aligned}
    &{\bf Q}_{e,k} = \tfrac{P}{\sigma_e^2}{\bf P}_{e,k}^{-1}{\bf H}_{e,k}, \quad {\bf Q}_{b,k} = \tfrac{P}{\sigma_b^2}{\bf P}_{b,k}^{-1}{\bf H}_{b,k},\\
    & {\bf P}_{e,k} ={\bf I}_{N_e} + \tfrac{P}{\sigma_e^2} {\bf H}_{e,k}{\bf W}_k{\bf W}_k^H{\bf H}_{e,k}^H, \\
    &{\bf P}_{b,k} ={\bf I}_{N_b} + \tfrac{P}{\sigma_b^2} {\bf H}_{b,k}{\bf W}_k{\bf W}_k^H{\bf H}_{b,k}^H,\\
    &{\bf M}_{i_1,k}={\bf W}_k{\bf W}_k^H{\bf H}_{a{i_1}}^H,\quad{\bf M}_{i_2,k}={\bf W}_k{\bf W}_k^H{\bf H}_{a{i_2}}^H.
\end{aligned}    
\end{equation}

\subsubsection{Determination of the Step Size}
Utilizing the Armijo linear search strategy \cite{boyd2004convex}, we dynamically adjust the step size during updates. This adaptive approach aligns the search step size with variations, preserving the property of non-increasing objective function values and enhancing algorithmic convergence speed. In particular, the chosen step size \( \alpha^q \) has to satisfy,
\begin{equation}
\mathcal{L}({\bf \Upsilon}^{q+1}) -\mathcal{L}({\bf \Upsilon}^{q})\le -\frac{1}{2}\alpha^q \|\text{grad}_{\mathcal{M}} \mathcal{L}({\bf \Upsilon}^q)\|_F^2. \label{stepsize}
\end{equation}

\subsubsection{Update and Retraction}
With the descent direction obtained from (\ref{calculReg}) and the step size determined by (\ref{stepsize}), the update in \( \mathcal{T}_{\bf \Upsilon} \mathcal{M} \) is given by,
\begin{equation}
    {  \bf  \Upsilon}^T = {\bf \Upsilon}^{q} - \alpha^q \text{grad}_{\mathcal{M}} f({\bf \Upsilon}^q).\label{Updateupsi}
\end{equation}
Since PRM \( \mathcal{M} \) is nonlinear, updates performed in the tangent space may result in \( {\bf \hat \Upsilon} \) falling outside of \( \mathcal{M} \). To address this, a retraction is applied to map the updated point from \( \mathcal{T}_{{\bf \Upsilon}^q} \mathcal{M} \) back onto the manifold \( \mathcal{M} \), as given by,
\begin{equation}
   {\bf \Upsilon}^{q+1}=  ({\bf \hat W}^{T}\oslash{\|{\bf \hat W}^{T}\|_F^2}) \oplus ( {\bm \phi}_1^T \oslash |{\bm \phi}_1^T|)\oplus ( {\bm \phi}_2^T \oslash |{\bm \phi}_2^T|), \label{useforretrac}
\end{equation}
where \textcolor{blue}{$\oslash$} and $(\cdot)^T$ denote the Hadamard division and transpose operators, respectively.

\begin{algorithm}
	\floatname{algorithm}{Algorithm}
	\renewcommand{\algorithmicrequire}{\textbf{Input:}}
	\renewcommand{\algorithmicensure}{\textbf{Output:}}
	\caption{: The PRGD algorithm to the problem (\ref{consr}).}
	\label{alg:1}
	\begin{algorithmic}[1]
		\STATE {\textbf {Initialize}}
            ${\bf \hat W}^q$, ${\bm \phi}_1^q$, ${\bm \phi}_2^q$, and set $q=0$.\\  
            \STATE Combine ${\bm \Upsilon}^q={\bf \hat W}^q \oplus  {\bm \phi}^q_1\oplus {\bm \phi}^q_2 $
            \STATE {\textbf {Reapeat}}
            \STATE \quad Update $\text{grad}_{\mathcal{M}} f({\bf \Upsilon}^q)$ by (\ref{calculReg});
            \STATE \quad Update \( \alpha^q \) by (\ref{stepsize});
            \STATE \quad Update ${\bf \Upsilon}^{q+1}$ by (\ref{Updateupsi}) and (\ref{useforretrac});
            \STATE \quad $q \leftarrow q+1$;
		\STATE {\textbf {Until} some stopping criterion is satisfied.}\\
            \STATE Decompose ${\bf \Upsilon}^{q}$ into ${\bf \hat W}^q$, ${\bm \phi}_1^q$, and ${\bm \phi}_2^q$;\\
             \ENSURE
                ${\bf \hat W}^q$, ${\bm \phi}_1^q$, and ${\bm \phi}_2^q$.\\
	\end{algorithmic}%
\end{algorithm}

In summary, the PRGD algorithm for solving (\ref{consr}) is presented in Algorithm \ref{alg:1}.

\subsection{Analysis of Complexity and Convergence}
In analyzing the computational complexity of the PRGD algorithm in Algorithm \ref{alg:1}, we focus on the primary contributors: \(M\), \(N_{i_1}\), and \(N_{i_2}\), assuming that \(N_s, N_e, N_b \ll M, N_{i_1}, N_{i_2}\). The primary complexity of the algorithm arises from the computation of the Riemannian gradient \(\text{grad}_{\mathcal{M}} f({\bf \Upsilon})\) as given in (\ref{calculReg}), which is of the order \(\mathcal{O}(K(M^2 \max \{N_{i_1}, N_{i_2}\} + \max \{N_{i_1}, N_{i_2}\} M^2))\). Assuming a total of \(Q\) iterations of the proposed algorithm, the overall computational complexity is \(\mathcal{O}(Q K (M^2 \max \{N_{i_1}, N_{i_2}\} + \max \{N_{i_1}, N_{i_2}\} M^2))\). For the convergence, we show that every limit point of Algorithm \ref{alg:1} is a stationary point of problem (\ref{consr}), as detailed in the Appendix.

\section{Numerical Results}
In this section, we compare the proposed method with the following benchmarks: 1) \textbf{GD-IRS} \cite{fukuda2011convergence}: gradient descent method over the Euclidean space to solve the problem, followed by direct projection after each iteration to satisfy the constraints; 2) \textbf{AOM-IRS} \cite{zhang2021multiple}: AO-based method that decouples the problem and applies the RGD algorithm to solve each subproblem; 3) \textbf{DD-IRS} \cite{rafieifar2023secure}: MIMO-OFDM system with uncoupled distributed double-IRS solved using the proposed method; 4) \textbf{SBob-IRS} \cite{jiang2021joint}: MIMO-OFDM system with a single IRS 1 deployed near Bob, whose number of REs is set to $N_{\text{sg}}=N_{i_1}+N_{i_2}$; 5) \textbf{SAlice-IRS} \cite{jiang2021joint}: MIMO-OFDM system with a single IRS 2 deployed near Alice, also using $N_{\text{sg}}$ REs; and 6) \textbf{R-IRS}: MIMO-OFDM system with cooperative double-IRS using randomly generated reflection matrices.

In our simulations, unless otherwise specified, the system parameters are set as follows: $M = 16$, $N = N_{i_1} = N_{i_2} = 48$, $N_s = N_b = N_e = 2$, $K = 10$, $P = 0\text{dB}$, and $\sigma_b^2 = \sigma_e^2 = -80\text{dBm}$. Consistent with the fading environment analyzed in \cite{rafieifar2023secure}, all wireless links experience both large-scale and small-scale fading. The small-scale fading follows a complex Gaussian distribution with zero mean and unit variance \cite{rafieifar2023secure}. The large-scale fading is modeled as $\text{PL} = \text{PL}_0 - 10\zeta \log_{10}(d/d_0)\text{dB}$, where $\text{PL}_0$ denotes the path loss at the reference distance $d_0$, $\zeta$ is the path loss exponent, and $d$ is the distance between the transmitter and the receiver. We set $\text{PL}_0 = -30\text{dB}$ and $d_0 = 1\text{m}$. The path loss exponent $\zeta$ is set to 2.5 for the Alice–IRS 1 and IRS 2–Bob/Eve links, 1.1 for the IRS 1–IRS 2 link, and 3.0 for all other links. The coordinates (in meters) of Alice, IRS 1, IRS 2, Eve, and Bob are $(0,0)$, $(10,10)$, $(50,10)$, $(40,0)$, and $(60,0)$, respectively. All simulation results are averaged over 100 independent random channel realizations.

\begin{figure*}[t]
    \centering
    \begin{minipage}[t]{0.32\textwidth}
        \centering
        \includegraphics[width=\linewidth]{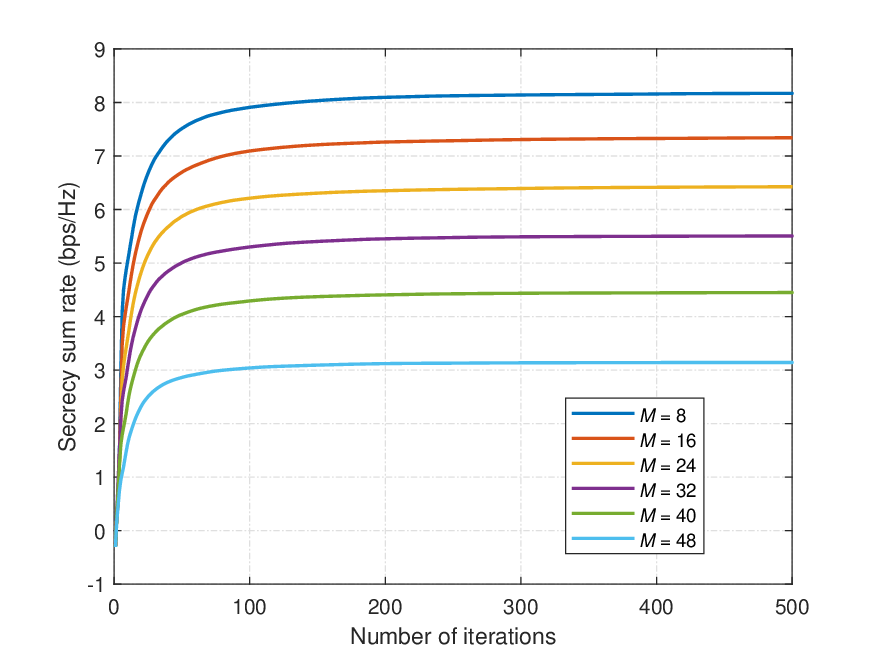}
        \caption{\footnotesize Analysis of the convergence performance.}
        \label{Convergcenoverall}
    \end{minipage}\hfill
    \begin{minipage}[t]{0.32\textwidth}
        \centering
        \includegraphics[width=\linewidth]{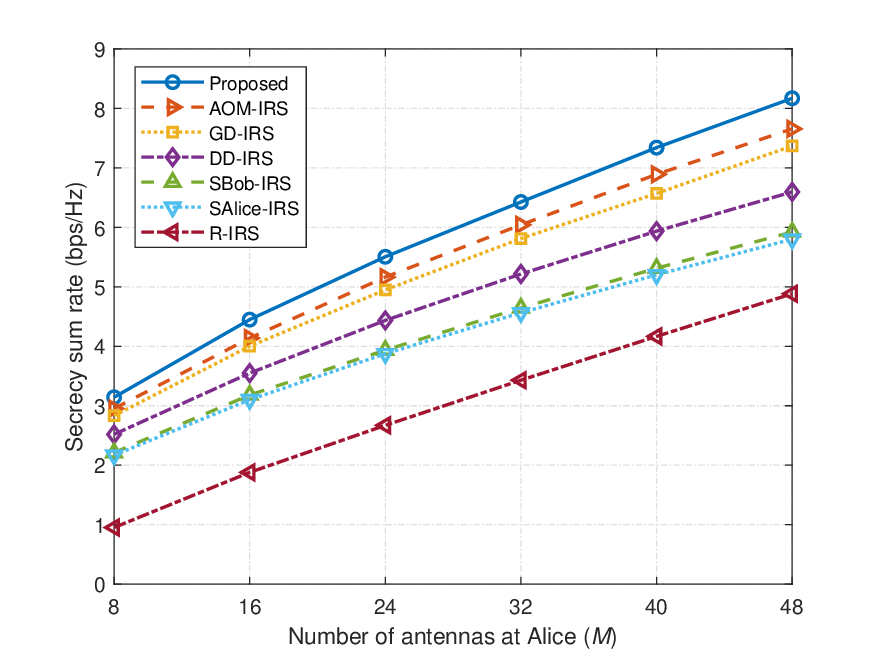}
        \caption{\footnotesize SSR versus the number of antennas.}
        \label{DifferentElement}
    \end{minipage}\hfill
    \begin{minipage}[t]{0.32\textwidth}
        \centering
        \includegraphics[width=\linewidth]{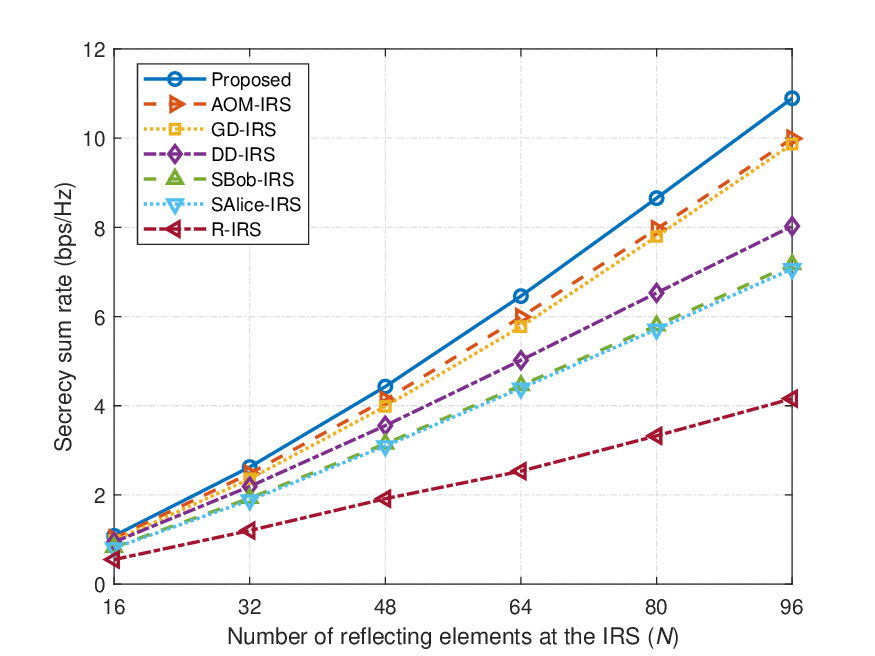}
        \caption{\footnotesize SSR versus the number of REs.}
        \label{DifferentDistance}
    \end{minipage}
    \vspace{-0.5cm}
\end{figure*}

\begin{figure*}[t]
    \centering
    \begin{minipage}[t]{0.32\textwidth}
        \centering
        \includegraphics[width=\linewidth]{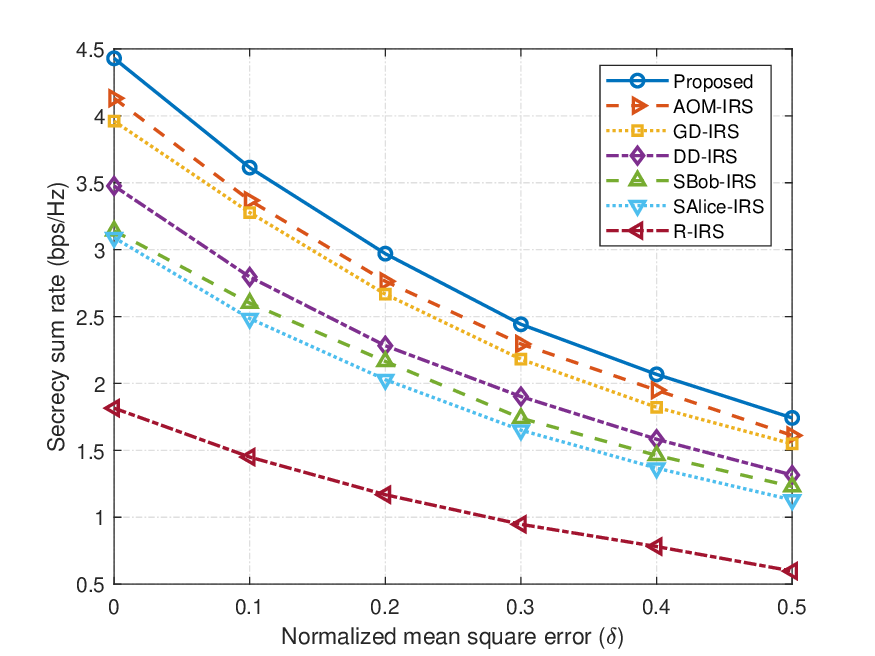}
        \caption{\footnotesize {SSR versus the NMSE.}}
        \label{NMSE}
    \end{minipage}\hfill
    \begin{minipage}[t]{0.32\textwidth}
        \centering
        \includegraphics[width=\linewidth]{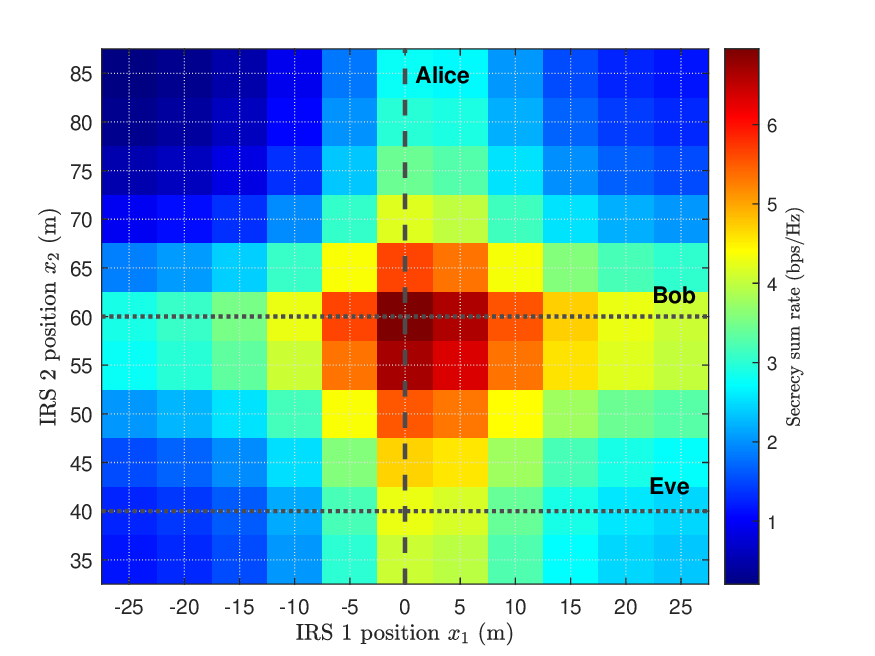}
        \caption{\footnotesize {SSR versus the position of IRSs.}}
        \label{Diffposi}
    \end{minipage}\hfill
    \begin{minipage}[t]{0.32\textwidth}
        \centering
        \includegraphics[width=\linewidth]{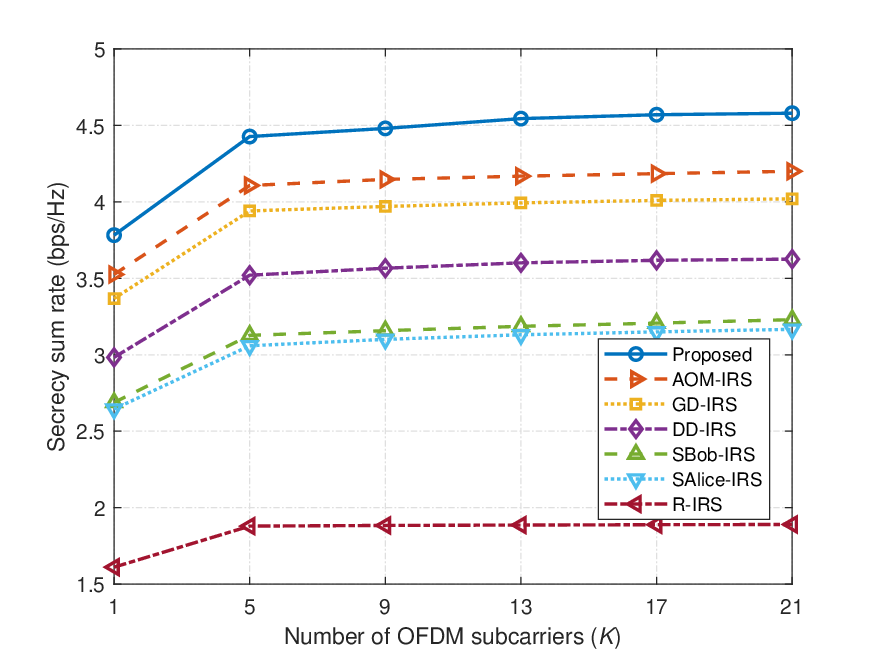}
        \caption{\footnotesize {SSR versus the number of subcarriers.}}
        \label{Differentcar}
    \end{minipage}
    \vspace{-0.5cm}
\end{figure*}

Fig. \ref{Convergcenoverall} shows the convergence of the proposed PRGD algorithm for different numbers of antennas at Alice ($M$). The SSR improves with larger $M$ due to increased spatial degrees of freedom. In all cases, the algorithm converges monotonically within 200 iterations, confirming its fast convergence and high efficiency.

Fig. \ref{DifferentElement} plots the SSR versus the number of transmit antennas at Alice ($M$). The proposed scheme consistently achieves higher secrecy rates as $M$ increases. At $M=16$, it attains 4.45 bps/Hz, improving over DD-IRS (3.55 bps/Hz) by 20.2\%, over SBob-IRS/SAlice-IRS (3.18 bps/Hz) by 28.5\%, and over R-IRS (1.88 bps/Hz) by 57.8\%. It also surpasses AOM-IRS (4.14 bps/Hz) and GD-IRS (4.00 bps/Hz) by 7.0\% and 10.1\%, respectively, confirming the benefit of the cooperative double-IRS architecture and the effectiveness of the product Riemannian optimization framework.

Fig. \ref{DifferentDistance} shows the SSR versus the number of REs ($N$). All schemes improve with larger $N$ due to stronger reflection capability. At $N=64$, the proposed cooperative double-IRS scheme attains 6.46 bps/Hz, exceeding AOM-IRS (5.99 bps/Hz), GD-IRS (5.77 bps/Hz), DD-IRS (5.02 bps/Hz), SBob-IRS (4.45 bps/Hz), SAlice-IRS (4.39 bps/Hz), and R-IRS (2.53 bps/Hz), with relative gains of 7.3\%, 10.7\%, 22.3\%, 31.1\%, 32.0\%, and 60.8\%, respectively. These results clearly demonstrate the superiority of the proposed scheme for enhancing PLS.

To relax the perfect CSI assumption, we evaluate the robustness of our scheme against CEEs. For any true channel $\mathbf H$, we model channel estimates as $\hat{\mathbf H}=\mathbf H-\mathbf E$ with i.i.d. $\mathbf E_{ij}\sim\mathcal{CN}(0,\sigma_E^2)$, and quantify CSI quality by the normalized mean square error (NMSE) $\delta=\mathbb{E}[\|\mathbf E\|_F^2]/\mathbb{E}[\|\mathbf H\|_F^2]$. Accordingly, $\sigma_E^2=\delta\mathbb{E}[\|\mathbf H\|_F^2]/M_e$, where $M_e$ denotes the number of elements in $\mathbf H$. The same model is applied to all channels and subcarriers. Fig. \ref{NMSE} plots the SSR versus the NMSE. As expected, the performance of all schemes degrades as the channel estimation becomes less accurate. However, the proposed cooperative double-IRS scheme consistently maintains a significant performance advantage over all benchmarks across the entire range of channel uncertainty. These results confirm that the architectural and algorithmic gains of our proposed solution are robust and hold in more practical scenarios with imperfect CSI.

Fig. \ref{Diffposi} investigates the optimal IRS deployment by showing the SSR versus the horizontal positions of IRS 1 ($x_1$) and IRS 2 ($x_2$), while keeping their vertical positions fixed. The results clearly indicate that the maximum SSR occurs in a compact region centered at $(x_1, x_2) \approx (0, 60)$ m, consistent with the classical guideline of placing one IRS near Alice and the other near Bob. Any deviation, such as moving IRS 2 toward Eve, is detrimental since the degradation of the legitimate channel outweighs any interference introduced to the eavesdropper. Thus, the presence of an eavesdropper does not alter the fundamental deployment rule but makes adherence to it more critical by substantially shrinking the high-performance region.

Fig. \ref{Differentcar} shows the SSR versus the number of OFDM subcarriers ($K$) under a fixed total power budget with frequency-flat IRS phases shared across all subcarriers. As $K$ increases, the SSR of all schemes improves. For example, the proposed design grows from about 3.8 bps/Hz at $K=1$ to about 4.6 bps/Hz at $K=21$, even though power is split across subcarriers and the IRS phases are common to all of them. This demonstrates the intrinsic benefit of OFDM, since accumulating secrecy over multiple subcarriers together with per-subcarrier beamforming achieves higher secrecy rates than single-carrier transmission. Across all values of $K$, the proposed scheme attains the highest SSR and maintains a growing advantage over all benchmarks, confirming its scalability in wideband IRS-assisted MIMO-OFDM systems.

\section{Conclusion}
This paper studied PLS enhancement in cooperative double-IRS–assisted MIMO-OFDM systems. The SSR maximization problem was formulated as a non-convex optimization and reformulated over a product Riemannian manifold, leading to an unconstrained problem. An efficient PRGD algorithm with guaranteed convergence was then developed. Simulations show that the proposed scheme achieves significant secrecy gains over existing benchmarks.

\section*{Appendix}
To establish the proof, we first show that the objective function in (\ref{consr}) decreases sufficiently at each iteration. This is ensured by carefully choosing the step size in (\ref{stepsize}), i.e.,
\begin{equation}
\begin{aligned}
\mathcal{L}(\bm{\Upsilon}^{q}) - \mathcal{L}(\bm{\Upsilon}^{q+1}) \ge c_{\text{dec}} \| \operatorname{grad}_{\mathcal{M}} \mathcal{L}(\bm{\Upsilon}^q) \|_F^2,\label{therem4proof}
\end{aligned}
\end{equation}
where $c_{\text{dec}} =\tfrac{1}{2}\alpha^k >0$. We can now complete the proof. The proof is based on a standard telescoping sum argument. The desired inequality for all $q=0,1,\dots,Q-1$ is obtained as,
\begin{subequations}
    \begin{align}
      {  \mathcal{L}(\bm{\Upsilon}^0) - }&{ \mathcal{L}^{\text{low}} \ge \mathcal{L}(\bm{\Upsilon}^0) - \mathcal{L}(\bm{\Upsilon}^Q) }\\ & { = \textstyle\sum_{q=0}^{Q-1} \mathcal{L}(\bm{\Upsilon}^q) - \mathcal{L}(\bm{\Upsilon}^{q+1})} \\&{ \ge Q c_{\text{dec}} \min _{q=0,1,\dots,Q-1}\| \operatorname{grad}_{\mathcal{M}}\mathcal{L}(\bm{\Upsilon}^q) \|_F^2}
    \end{align}
\end{subequations}
where $\mathcal{L}^{\text{low}}$ is the lower bound value for the objective function in (\ref{consr}). To get the limit statement, observe that $\mathcal{L}(\bm{\Upsilon}^{q+1}) \le \mathcal{L}(\bm{\Upsilon}^q)$ for all $q$ by (\ref{therem4proof}). Then, taking $Q$ to infinity we have,
\begin{equation}
  {  \mathcal{L}(\bm{\Upsilon}^0) -  \mathcal{L}^{\text{low}} \ge \textstyle\sum_{q=0}^{\infty} \mathcal{L}(\bm{\Upsilon}^q) - \mathcal{L}(\bm{\Upsilon}^{q+1}),}
\end{equation}
where the right-hand side is a series of nonnegative numbers. The bound implies that the summands converge to zero, thus,
\begin{equation}
\begin{aligned}
  {   0} &{ = \lim_{q \to \infty} \mathcal{L}(\bm{\Upsilon}^q) - \mathcal{L}(\bm{\Upsilon}^{q+1}) }\\ &{ \ge c_{\text{dec}} \lim_{q \to \infty}\| \operatorname{grad}_{\mathcal{M}} \mathcal{L}(\bm{\Upsilon}^q) \|_F^2,  }  \end{aligned}
\end{equation}
which confirms that $\|\operatorname{grad}_{\mathcal{M}} \mathcal{L}(\bm{\Upsilon}^q) \|_F\to0$. Now, let ${\bm{\Upsilon}}^*$ be a limit point of the sequence of iterates. By definition, there exists a subsequence of iterates
${\bm{\Upsilon}}^{(0)},{\bm{\Upsilon}}^{(1)},{\bm{\Upsilon}}^{(2)},\dots$ which converges to ${\bm{\Upsilon}}^*$. Then, since the norm of the gradient of $ \mathcal{L}(\bm{\Upsilon})$ is a continuous function, it commutes with the limit and we have,
\begin{equation}
\begin{aligned}
 {   0} &{ = \lim_{q \to \infty}\| \operatorname{grad}_{\mathcal{M}} \mathcal{L}(\bm{\Upsilon}^q) \|_F^2 = \lim_{q \to \infty}\| \operatorname{grad}_{\mathcal{M}} \mathcal{L}(\bm{\Upsilon}^{(q)}) \|_F^2} \\ &{  = \|\operatorname{grad}_{\mathcal{M}} \mathcal{L}(\lim_{q \to \infty}( \bm{\Upsilon}^{(q)})) \|_F^2 = \|\operatorname{grad}_{\mathcal{M}} \mathcal{L}({\bm{\Upsilon}}^*) \|_F^2,}
\end{aligned}
\end{equation}
showing that all limit points generated by the sequence are stationary points. This completes the proof.$\hfill\blacksquare$

\bibliographystyle{IEEEtran}
\bibliography{strings}

\vfill

\end{document}